\documentclass[aps,twocolumn]{revtex4-2}
\usepackage{amssymb}
\usepackage{graphicx}
\usepackage{dcolumn}
\usepackage{bm}
\usepackage{amsmath}
\usepackage{soul,color}
\usepackage{textcomp}
\usepackage{times}
\usepackage[colorlinks,linkcolor=red,citecolor=blue]{hyperref}
\graphicspath{{figures/}}

\begin{document}

\title{Spin-orbit coupling induced geometric squeezing in rotating Bose-Einstein condensates}

\author{Fei Zhu$^{1}$}
\author{Chunxia Guo$^{1}$}
\author{Rui Zhang$^{1}$}
\author{Lianghui Huang$^{2,4}$}
\email{huanglh06@sxu.edu.cn}
\author{Ren Zhang$^{3,4}$}
\email{renzhang@xjtu.edu.cn}
\author{Li Chen$^{1}$}
\email{lchen@sxu.edu.cn}

\affiliation{
$^1${Institute of Theoretical Physics, State Key Laboratory of Quantum Optics and Quantum Optics Devices, Shanxi University, Taiyuan 030006, China}\\
$^2${State Key Laboratory of Quantum Optics Technologies and Devices,
Institute of Opto-electronics, Collaborative Innovation Center of Extreme Optics,
Shanxi University, Taiyuan, Shanxi 030006, China}\\
$^3${MOE Key Laboratory for Nonequilibrium Synthesis and Modulation of Condensed Matter, Shaanxi Province Key Laboratory of Quantum Information and Quantum Optoelectronic Devices, School of Physics, Xi’an Jiaotong University, Xi’an 710049, China}\\
$^4${Hefei National Laboratory, Hefei, 230088, China}\\
}

\begin{abstract}
Squeezed states play a key role in diverse frontiers of quantum physics. Geometrically squeezed states, a squeezed state 
in the orbital phase space of rotating Bose-Einstein condensates (BEC), have been conventionally generated by anisotropic trapping potentials. In this work, we propose a different route to generate geometric squeezing via spin-orbit coupling (SOC) in a pseudospin-1/2 BEC. We show that the SOC enables effective two-phonon transitions within the lowest Landau level via virtual spin-flip processes, leading to exponential squeezing dynamics in both spin components. Furthermore, by applying a $\pi/2$ spin rotation, the two spin channels can be coherently coupled to produce two-mode geometric squeezing. We also investigate the influence of interatomic interactions on squeezing performance and identify parameters where robust squeezing can be achieved. Our work provides a viable pathway to realize and manipulate geometric squeezing in spinor quantum gases.
\end{abstract}

\maketitle

\section{Introduction}
\label{sec:introduction}

Squeezed states are characterized by the reduction of quantum fluctuations in one phase-space quadrature below the standard quantum limit (SQL), at the expense of increased quantum noise in the conjugate variable ~\cite{scully1997,walls2008}. Intuitively, squeezed states are defined by an elliptical uncertainty distribution in phase space, in contrast to the circular distribution of coherent states. The squeezed states serve as critical resources in various topics of quantum information science, such as quantum-enhanced sensing~\cite{braunstein2005,degen2017}, communication~\cite{braunstein2005,weedbrook2012}, and computation~\cite{braunstein2005,weedbrook2012}. Since the pioneering experimental demonstrations in optical parametric oscillators (OPOs)~\cite{kimble1986,kimble1987,slusher1985}, squeezed states have been extensively studied and observed across a wide variety of quantum platforms, including optomechanical resonators~\cite{aspelmeyer2014,safavi2013,wollman2015}, trapped ions~\cite{kienzler2015,fluhmann2019}, atomic ensembles~\cite{julsgaard2001,appel2008}, and superconducting circuits~\cite{eichler2011,menzel2012}. Notably, squeezed light has enabled gravitational-wave detectors such as LIGO to surpass the SQL~\cite{ligo2011}, and has been applied to quantum teleportation~\cite{furusawa1998} and secure communication protocols~\cite{hillery2000,grosshans2003}.

The \textit{geometric squeezing} refers to the squeezed states %existing 
in orbital phase spaces of rotating Bose-Einstein condensates (BEC), pioneered by the seminal experiment by Fletcher \textit{et al.}~\cite{fletcher2021}. When the rotation frequency $\Omega$ of the trapping potential approaches the radial confinement frequency $\omega$, the centrifugal force nearly cancels the external confining potential, leading to a massive energy degeneracy in the guiding-center mode. This degeneracy corresponds to the formation of Landau levels, analogous to the %behavior of 
charged particles in a magnetic field.
In this regime, an additional weak anisotropic potential ($\propto x^2-y^2$) induces two-phonon transitions within the lowest Landau level (LLL), closely mirroring the two-photon processes in OPOs~\cite{kimble1986,kimble1987}. This mechanism enables the generation of single-mode geometric squeezed states, where the quadrature uncertainty is exponentially compressed in one direction of the guiding-center phase space. Subsequent investigations have explored the dynamic control of rotation frequency and anisotropic strength to optimize the production of both single-mode~\cite{sharma2022,chen2025dyn,crepel2024} and two-mode geometric squeezing~\cite{chen2025}, as well as the rich interplay between geometric squeezing and interatomic interactions, leading to exotic phenomena such as bosonic quantum Hall states and vortex crystals~\cite{fetter2009,cooper2008,viefers2008,regnault2003,furukawa2012,regnault2013,mukherjee2022}.

In addition to the external orbital degrees of freedom, ultracold atoms possess rich and controllable internal states~\cite{kawaguchi2012}, i.e., the pseudospin degrees of freedom. By engineering transitions between different spin states and exploiting state-dependent interactions, spinor BECs have emerged as a versatile platform for exploring exotic quantum phenomena, including magnetic phases~\cite{stenger1998,chang2004,schmaljohann2004,kronjager2005}, topological excitations~\cite{sadler2006,isoshima2000,mizushima2002,saito2006}, and quantum turbulence~\cite{seo2016,weiler2008}. In spinor quantum gases, one the most significant breakthroughs has been the experimental realization of spin-orbit coupling (SOC)~\cite{lin2011,Wang2012,Cheuk2012}.
Typically, Raman-induced SOC of the form $\sigma_x p_x$ (where $\sigma_x$ is the Pauli spin operator and $p_x$ is the momentum operator along the $x$-direction), coherently couples the atom's internal spin to its external momentum.
The SOC gives rise to a variety of novel phenomena~\cite{galitski2013,goldman2014,zhai2015,zhangyi2018}, such as striped supersolid-like phases~\cite{Wang2010,Wu2011,ho2011,li2012,zheng2013,li2017,demarco2015,lchenpu2016,sun2015}, topological phase transitions~\cite{wu2016,huang2016,jotzu2014,aidelsburger2013}, exotic vortex configurations~\cite{hu2012,lchenpu2020}, and spin-squeezed or spin-entangled states~\cite{lchenpu2020ss,zhu2025}. 
However, previous studies~\cite{fletcher2021,sharma2022,chen2025dyn,crepel2024,chen2025} on geometric squeezing have been confined to scalar BECs, where all atoms occupy a single internal state with the spin degrees of freedom being frozen. This raises a natural question: can SOC induce geometric squeezing in spinor BECs?

In this paper, we demonstrate that the Raman-induced SOC is capable of generating single-mode geometric squeezing in each spin component. Our scheme operates through a mechanism fundamentally distinct from the traditional approach based on anisotropic trapping potentials. In contrast to the anisotropic potential ($\propto x^2-y^2$) that is inherently quadratic, the SOC term ($\propto \sigma_x p_x$) is merely linear in the momentum operator. Nevertheless, we show that second-order virtual processes within the LLL manifold result in an effective quadratic Hamiltonian, which serves as the key driver for geometric squeezing. Building upon this, we further demonstrate that a $\pi/2$ spin rotation can coherently couple the two spin components, leading to the two-mode geometric squeezing. Additionally, the influence of interatomic interactions on the squeezing dynamics is also discussed.

The rest of the paper is organized as follows. In Sec.~\ref{sec:aniso_squeezing}, we review the geometric squeezing induced by anisotropic trapping in scalar BECs. In Sec.~\ref{sec:soc_squeezing}, we present the mechanism of SOC-induced geometric squeezing in a spin-1/2 BEC, and provide numerical validation of single-mode squeezing dynamics. In Sec.~\ref{sec:two_mode_squeezing}, we discuss the generation of two-mode geometric squeezing via spin rotation and characterize the joint noise properties. In Sec.~\ref{sec:discussion}, we analyze the effects of interatomic interactions on the squeezing performance. In Sec.~\ref{sec:observation}, we discuss the experimental realization and an observation scheme for the guiding-center quadrature variances. Finally, a summary of our findings is given in Sec.~\ref{sec:conclusion}. 

\section{Anisotropic Trap Induced Geometric Squeezing}
\label{sec:aniso_squeezing}

We first review how geometric squeezing emerges in a scalar rotating BEC subject to a weak trap anisotropy~\cite{fletcher2021,sharma2022,chen2025dyn,crepel2024,chen2025}.
The central insight is that as the rotation frequency $\Omega$ approaches the harmonic trapping frequency $\omega$, the condensate naturally decomposes into two independent sectors: a high-energy \textit{cyclotron} mode and a near-degenerate \textit{guiding-center} mode.
Then, a weak harmonic confinement leads to an exponential deformation of the quantum-noise ellipse in the guiding-center phase space.

The single-particle Hamiltonian in the rotating frame (setting $\hbar=1$) reads
\begin{eqnarray}\label{eq:h0_rot}
	h_0 = \frac{\mathbf{p}^2}{2m} + V_0(\mathbf{r}) - \Omega L_z,
\end{eqnarray}
where $\mathbf{r}=(x,y)^{\mathsf T}$ and $\mathbf{p}=(p_x,p_y)^{\mathsf T}$ denote the position and momentum operators, $L_z = x p_y - y p_x$ is the $z$-component of angular momentum, and $V_0(\mathbf{r}) = \tfrac12 m\omega^2 r^2$ is an isotropic harmonic trap. 
Near the critical rotation $\Omega\approx\omega$, the phase-space structure naturally separates into two independent sectors, i.e., the cyclotron and guiding-center mode. For the cyclotron mode,
\begin{eqnarray}\label{eq:a_mode}
	a = \frac{\xi + i \eta}{\sqrt{2}\, l_B}, \qquad
	a^\dagger = \frac{\xi - i \eta}{\sqrt{2}\, l_B},
\end{eqnarray}
where
\begin{eqnarray}
	\xi = \frac{x}{2} - \frac{p_y}{2 m \omega}, \qquad
	\eta = \frac{y}{2} + \frac{p_x}{2 m \omega};
\end{eqnarray}
while for the guiding-center mode (LLL geometric degree of freedom),
\begin{eqnarray}
	b = \frac{X - i Y}{\sqrt{2}\, l_B}, \qquad
	b^\dagger = \frac{X + i Y}{\sqrt{2}\, l_B},
\end{eqnarray}
with the conjugate quadratures
\begin{eqnarray}\label{eq:b_mode}
	X = \frac{x}{2} + \frac{p_y}{2 m \omega}, \qquad
	Y = \frac{y}{2} - \frac{p_x}{2 m \omega}.
\end{eqnarray}
Here, $l_B = 1/\sqrt{2m\omega}$ is the magnetic length, which sets the characteristic width of LLL wave packets~\cite{fetter2009,viefers2008}. The two sectors are decoupled: each pair $(\xi,\eta)$ and $(X,Y)$ spans a canonically conjugate phase space, while operators from different sectors commute, i.e.,
\begin{equation}
	\begin{aligned}
		\left[a,a^\dagger\right] &= \left[b,b^\dagger\right]=1, \\
		\left[\xi,\eta\right] &= -\left[X,Y\right] = i l_B^2, \\
		\left[a,b\right] &= \left[a,b^\dagger\right]=0.
	\end{aligned}
\end{equation}

Expressed in this modal basis, the Hamiltonian decomposes into two independent harmonic oscillators, i.e.,
\begin{eqnarray}\label{eq:h0_ab}
	h_0
	&=& m \omega\Big[\omega_+\left(\xi^2+\eta^2\right)
	+\omega_-\left(X^2+Y^2\right)\Big] \nonumber \\
	&=&\omega_+\left(a^\dagger a+\frac12\right)
	+\omega_-\left(b^\dagger b+\frac12\right),\label{eq:h0}
\end{eqnarray}
where $\omega_\pm = \omega \pm \Omega$. The eigenbasis is labeled by $|n_{a},n_{b}\rangle$, with non-negative integers $n_{a}$ and $n_{b}$ counting excitations in the cyclotron and guiding-center modes. 
At the critical rotation frequency $\Omega = \omega$, the guiding-center frequency $\omega_-$ vanishes, rendering this sector completely degenerate. The spectrum collapses into Landau levels, with the LLL spanned by $|n_a=0,n_b\rangle$. 

The geometric squeezing can be dynamically realized by introducing a weak anisotropic trapping potential~\cite{fletcher2021}
\begin{equation}
V(\mathbf{r}) = V_0({\bf r}) + \frac{\varepsilon m \omega^2}{2} (x^2-y^2),
\label{eq:V_aniso}
\end{equation}
where $\varepsilon$ is a small dimensionless parameter. The anisotropy breaks the rotational symmetry, inducing quadratic coupling between the modes. To expose the resulting squeezing structure, one applies a unitary transformation $G = \exp(-i\kappa m \omega x y)$ with $\kappa = \varepsilon \omega/(2\Omega)$. Under $G$, the spatial coordinates remain unchanged while the momenta acquire linear corrections, $Gp_xG^\dagger = p_x + \kappa m\omega\, y$ and $Gp_yG^\dagger = p_y + \kappa m\omega\, x$; choosing $\kappa = \varepsilon\omega/(2\Omega)$ cancels the bare anisotropic potential exactly. Expressed in the $(a,b)$ ladder-operator basis and retaining terms to leading order in $\kappa\sim O(\varepsilon)$, the transformed Hamiltonian takes the form (see Appendix~\ref{app:eq9_validity} for a complete derivation)
\begin{eqnarray}
h_0 &=&\omega_+\left(a^{\dagger} a+\frac{1}{2}\right)+\omega_-\left(b^{\dagger} b+\frac{1}{2}\right) \nonumber \\
&&-\frac{ \zeta}{2}\left(a^{\dagger} a^{\dagger}+a a-b^{\dagger} b^{\dagger}-b b\right),
\label{eq:h0_anisotropic}
\end{eqnarray}
where $\zeta = \kappa \omega = \varepsilon \omega^2/(2\Omega)$ characterizes the strength of the two-phonon transition induced by the anistropic potential.

The crucial feature emerges at the critical rotation ($\Omega=\omega$). Here, $\omega_- = 0$ while the squeezing terms $\frac{\zeta}{2}(b^{\dagger} b^{\dagger}+b b)$ remain active. It 
drives the guiding-center mode toward a single-mode geometrically squeezed state (formally analogous to 
parametric down-conversion in quantum optics), i.e.,
\begin{equation}
	\exp(-i h_0 t)|0,0\rangle \approx |0,S(t)\rangle.
	\label{eq:squeezing_dy}
\end{equation}
The phase-space signature is an exponentially deforming noise ellipse in the $X$-$Y$ plane.
To make the relevant observables explicit, we introduce the rotated guiding-center quadrature $X(\theta) = X\cos\theta + Y\sin\theta$ and its variance
\begin{equation}
	\Delta^2(\theta,t) = \langle X(\theta)^2\rangle - \langle X(\theta)\rangle^2,
	\label{eq:Delta_theta}
\end{equation}
which characterizes the noise of the guiding-center mode along an arbitrary direction $\theta$ in the $X$-$Y$ phase space. The minimum and maximum quadrature variances $\Delta^2_{\min/\max}(t)$ are the extrema of $\Delta^2(\theta,t)$ over $\theta$, equivalently the two eigenvalues of the $X$-$Y$ covariance matrix.
For the squeezing dynamics governed by Eq.~(\ref{eq:h0_anisotropic}) at $\Omega=\omega$, they take the analytic form
\begin{equation}
\begin{aligned}
\Delta^2_\text{min}(t) &= \Delta^2_\text{SQL} e^{- 2\zeta t}, \\
\Delta^2_\text{max}(t) &= \Delta^2_\text{SQL} e^{2\zeta t},
\end{aligned}
\label{eq:Delta_min_max}
\end{equation}
where $\Delta^2_\text{SQL} = l_B^2/2$ is the SQL.
In contrast, the cyclotron mode remains unsqueezed due to the dominating term ${a}^{\dagger} {a}$ in Eq.~(\ref{eq:h0_anisotropic}) with strength $\omega_+ = 2\omega \gg \zeta$. 
This term induces rapid rotation of the $\xi$-$\eta$ phase space, thereby suppressing the formation of squeezing in the cyclotron mode.

\section{Spin-Orbit Coupling Induced Geometric Squeezing}
\label{sec:soc_squeezing}

\subsection{Effective Hamiltonian}

We now turn to an isotropic two-component BEC (pseudospin-1/2 system) subject to the Raman-induced SOC. 
We will demonstrate that the SOC can induce geometrically squeezed states within the framework of perturbation theory, a fundamentally distinct mechanism from the anisotropic potential approach described above.

The Hamiltonian of the spin-$1/2$ BEC with Raman-induced SOC is given by
\begin{equation}
	H = h_0\otimes\mathbb{I} - J\sigma_z + \alpha p_x\sigma_x,
	\label{eq:H_total}
\end{equation}
where $h_0$ is given in Eq.~(\ref{eq:h0}) and $\mathbb{I}$ is a $2\times 2$ identity matrix. The last term $\alpha p_x \sigma_x$ characterizes the SOC with $\sigma_{x,z}$ being the Pauli matrices, $p_x$ being the momentum operator, which can be expanded in terms of the ladder operators as
\begin{equation}
	p_x = \frac{i\sqrt{m\omega}}{2}\Big[(a^\dagger-a)+(b^\dagger-b)\Big],
	\label{eq:p_x}
\end{equation}
and $\alpha$ being the strength; $J$ being the Raman Rabi frequency. 
This form of SOC can typically be realized through a two-photon process in which two non-collinear, largely detuned laser beams illuminate the atoms~\cite{lin2011,galitski2013}.
Note that our SOC Hamiltonian differs from the commonly used form in the literature by a spin rotation, namely $\sigma_z \to \sigma_x$ and $\sigma_x \to -\sigma_z$. This rotation is merely for the convenience of discussion and does not lead to any additional physical consequences.
Since $p_x$ is linear in the ladder operators $a,b$ [Eq.~(\ref{eq:p_x})], the SOC term $\alpha p_x\sigma_x$ cannot directly generate the quadratic coupling $b^2+b^{\dagger 2}$ required for guiding-center squeezing, in contrast to the anisotropic potential which does so at first order [Eq.~(\ref{eq:h0_anisotropic})]. We will show that such a quadratic coupling is instead generated at second order in perturbation theory, through virtual inter-spin transitions within the LLL manifold.

\begin{figure}[t]
	\centering
	\includegraphics[width=\linewidth]{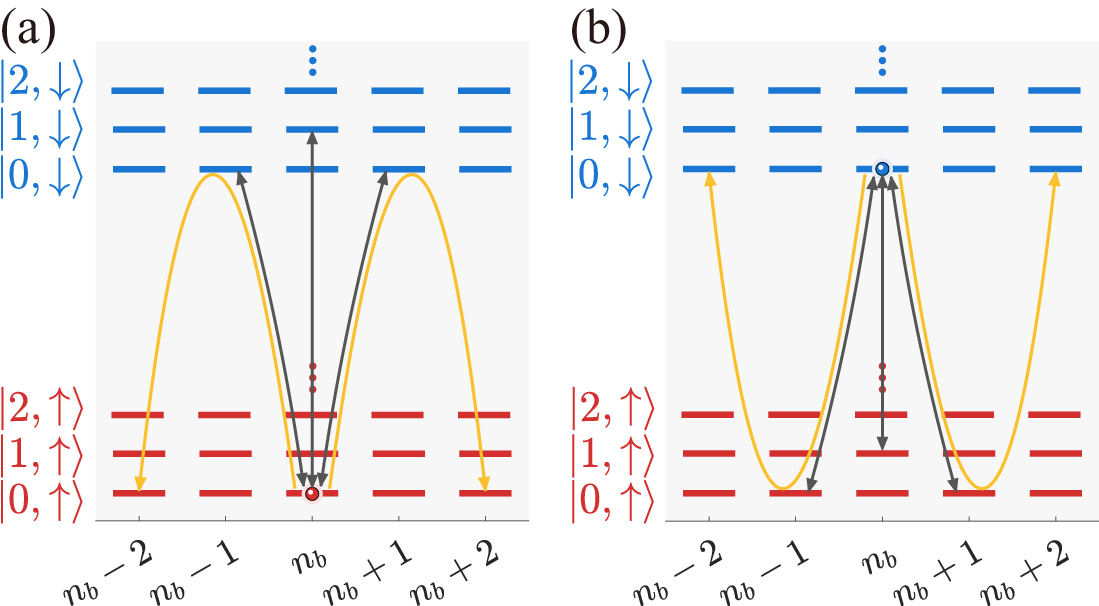}
	\caption{Landau level diagrams for the spin-orbit coupled rotating BEC with $\alpha = 0$. Panels (a) and (b) show the energy spectrum and virtual transition processes within the spin-up and spin-down subspaces, respectively. 
	The black double arrows and yellow single arrows, respectively, correspond to diagonal and off-diagonal corrections.
	}
	\label{fig:level}
\end{figure}

In the spin-1/2 BEC, the basis states are $|n_a,n_b,\sigma\rangle$, with the additional $\sigma = \{\uparrow, \downarrow\}$ denoting the spin index. In the absence of SOC ($\alpha=0$), we denote the Hamiltonian as $H_0 = H_{\alpha=0}=h_0\otimes\mathbb{I}-J\sigma_z$. The states $|n_a,n_b,\sigma\rangle$ are eigenstates of $H_0$. At $\Omega=\omega$, the Raman $J$ term simply breaks the spin degeneracy, resulting in a gap $2J$ between the Landau levels of different spin components, as shown in Fig.~\ref{fig:level}. In particular, within the LLL manifold $\{|0,n_b,\sigma\rangle\}$ one has $E_{0,\uparrow}=\omega-J$ and $E_{0,\downarrow}=\omega+J$.

Next, we turn on a weak SOC and treat
$H' \equiv \alpha p_x\sigma_x$ as a perturbation (i.e., $\alpha p_0 \ll J$ with $p_0\sim\sqrt{m\omega}$ being the oscillator momentum scale). 
Therefore, the effect of $H'$ on the LLL manifold is described by degenerate perturbation theory.
We first focus on the LLL of the spin-up component whose projector is $P_\uparrow=\sum_{n_b}|0,n_b,\uparrow\rangle\langle 0,n_b,\uparrow|$. Since $H'$ flips the spin, the first-order correction within this degenerate subspace vanishes, i.e., $P_\uparrow H' P_\uparrow=0$. Hence, the leading effect comes from the second-order correction, which yields the effective Hamiltonian
\begin{equation}
	H_{\rm eff}^{(\uparrow)} = P_\uparrow H' Q \frac{1}{E_{0,\uparrow}-H_0} Q H' P_\uparrow,
\end{equation}
where $Q=1-P_\uparrow$ is the projector onto the complementary subspace.

As illustrated in Fig.~\ref{fig:level}(a), there only exist two types of virtual processes contributing to $H_{\rm eff}^{(\uparrow)}$: (i) diagonal corrections, denoted by black arrows, arising from the vertical round-trip excitation $\left|0,n_b,\uparrow \right\rangle \to \left|1,n_b,\downarrow \right\rangle \to \left|0,n_b,\uparrow \right\rangle$ and the horizontal round-trip transition $\left|0,n_b,\uparrow \right\rangle \to \left|0,n_b\pm1,\downarrow \right\rangle \to \left|0,n_b,\uparrow \right\rangle$, which contribute effective $b^\dagger b$ terms;
(ii) off-diagonal processes, denoted by yellow arrows, from the two-step horizontal virtual transition 
$\left|0,n_b,\uparrow\right\rangle \to \left|0,n_b\pm1,\downarrow\right\rangle \to \left|0,n_b\pm2,\uparrow\right\rangle$,
which produces the $(b^\dagger)^2+b^2$ terms serving as the key ingredient for inducing single-mode geometric squeezing as we will demonstrate in Sec.~\ref{sec:numerical_results}.

%Carrying out detailed calculations, the effective Hamiltonian is found to be
A straightforward derivation yields the effective Hamiltonian
\begin{equation}
	H_{\rm eff}^{(\uparrow)} = -\zeta\, b^\dagger b + \frac{\zeta}{2}\left[(b^\dagger)^2+b^2\right] + \mathcal{E}_{\uparrow},
	\label{eq:Heff_up_final0}
\end{equation}where 
\begin{equation}
	\zeta \equiv \frac{\alpha^2 m\omega}{4J}
\end{equation}
is non-negative for $J > 0$, and $\mathcal{E}_{\uparrow}=\omega-J-\frac{\alpha^2 m\omega}{8J}-\frac{\alpha^2 m\omega}{8(\omega+J)}$ is a constant. 
On the other hand, for the spin-down LLL manifold, one can perform a similar analysis in the subspace $P_\downarrow=\sum_{n_b} \left|0,n_b,\downarrow \right\rangle \left\langle 0,n_b,\downarrow \right|$. 
All the contributing processes are visualized in Fig.~\ref{fig:level}(b), and the effective Hamiltonian turns out to be
\begin{equation}
	H_{\rm eff}^{(\downarrow)} = \zeta\, b^\dagger b - \frac{\zeta}{2}\left[(b^\dagger)^2+b^2\right] + \mathcal{E}_{\downarrow},
	\label{eq:Heff_down_final0}
\end{equation}with $\mathcal{E}_{\downarrow}=\omega+J+\frac{\alpha^2 m\omega}{8J}+\frac{\alpha^2 m\omega}{8(J-\omega)}$. 

For a single spin component ($\uparrow$ or $\downarrow$), the diagonal correction $\mp \zeta b^\dagger b$ in Eqs.~(\ref{eq:Heff_up_final0}) and (\ref{eq:Heff_down_final0}) can be removed by a small shift of the rotation frequency $\Omega$ away from the critical value $\omega$, according to Eq.~(\ref{eq:h0}). Specifically, we choose $\Omega = \omega-\zeta$ to cancel the diagonal term of the spin-up component, leading to 
\begin{equation}
	H_{\rm eff}^{(\uparrow)} = \frac{\zeta}{2}\left[(b^\dagger)^2+b^2\right].
	\label{eq:Heff_up_final}
\end{equation}
Similarly, we choose $\Omega=\omega+\zeta$ for the spin-down component yielding
\begin{equation}
	H_{\rm eff}^{(\downarrow)} = -\frac{\zeta}{2}\left[(b^\dagger)^2+b^2\right].
	\label{eq:Heff_down_final}
\end{equation}
Hence, the corresponding dynamics would lead to ideal single-mode geometric squeezed states ($|0,S(t),\uparrow\rangle$ or $|0,S(t),\downarrow\rangle$), with the quadrature variances in the $X_\sigma$-$Y_\sigma$ phase spaces following the exponential behaviors shown in Eq.~(\ref{eq:Delta_min_max}).

However, since the diagonal corrections for the spin-up and spin-down components have opposite signs, i.e., $\mp\zeta$, they cannot be simultaneously eliminated by a single shift of $\Omega$ away from the critical value $\omega$. 
Without any shift, i.e., at $\Omega=\omega$, both components retain their diagonal terms with symmetric strengths $\mp\zeta$. 
In this case, the dynamics exhibits sheared squeezing in both phase spaces, where the quadrature fluctuations follow
\begin{equation}
\begin{aligned}
\Delta^2_{\sigma,\min}(t) &= \Delta^2_{\rm SQL}\big(\sqrt{1+(\zeta t)^2}-\zeta t\big)^2, \\
\Delta^2_{\sigma,\max}(t) &= \Delta^2_{\rm SQL}\big(\sqrt{1+(\zeta t)^2}+\zeta t\big)^2.
\end{aligned}
\label{eq:shear_squeezing}
\end{equation}
At short times $\zeta t\ll 1$, $\Delta^2_{\sigma,\min/\max}(t)$ reproduces the exponential scaling $\propto e^{\mp 2\zeta t}$, whereas at longer times $t\sim 1/\zeta$, the squeezing dynamics deviates from the exponential scaling due to the phase-space rotation induced by the diagonal corrections.
Here and throughout the paper, $\Delta_\sigma^2(\theta,t)$ and $\Delta_{\sigma,\min/\max}^2(t)$ carry the same meaning as Eq.~(\ref{eq:Delta_min_max}), but in the spin-independent phase space $X_\sigma$-$Y_\sigma$. A detailed protocol for reconstructing these variances from spin-resolved real-space density images is presented in Sec.~\ref{sec:observation}.

\subsection{Numerical Results}
\label{sec:numerical_results}

Now, we present the numerical benchmark of the predictions mentioned in the previous subsection. Consider the following protocol: 
(i) Prepare the isotropic spinor BEC in the ground state without rotation and SOC;
(ii) Ramp up the rotation frequency $\Omega$ into the near-critical regime, during which both spin components remain isotropic;
(iii) At $t=0$, suddenly turn on the weak SOC (e.g., via Raman coupling). The system then evolves under the total Hamiltonian $H$ [Eq.~(\ref{eq:H_total})].
We simulate the squeezing dynamics by numerically solving the time-dependent Schr\"{o}dinger equation in the coordinate representation, i.e., $i\partial_t \boldsymbol{\Psi}(\mathbf{r},t) = H \boldsymbol{\Psi}(\mathbf{r},t)$, with $\boldsymbol{\Psi}(\mathbf{r},t) = (\psi_\uparrow(\mathbf{r},t), \psi_\downarrow(\mathbf{r},t))^\mathsf{T}$.

We take the BEC experiments with Raman-induced SOC as a reference~\cite{lin2011}, where two internal states are coupled using a pair of Raman lasers.
For two Raman beams of wavelength $\lambda$ intersecting at an angle $\theta$, the transferred momentum is $2 k_R$ with $k_R=(2\pi/\lambda)\sin(\theta/2)$, giving an SOC coefficient of order $\alpha\simeq k_R/m$; hence $\alpha$ can be tuned by changing the angle $\theta$.
The recoil energy $E_R=k_R^2/(2m)$ is typically in the kHz range, whereas radial trap frequencies are usually tens to about $100$ Hz, thus we take a Raman Rabi frequency $J=10\omega$ is a representative off-resonant parameter regime.
For the parameters used below, $\alpha=0.5\sqrt{\omega/m}$ gives $\alpha \sqrt{m\omega}\sim0.5\omega \ll J$.

\begin{figure}[t]
	\centering
	\includegraphics[width=\linewidth]{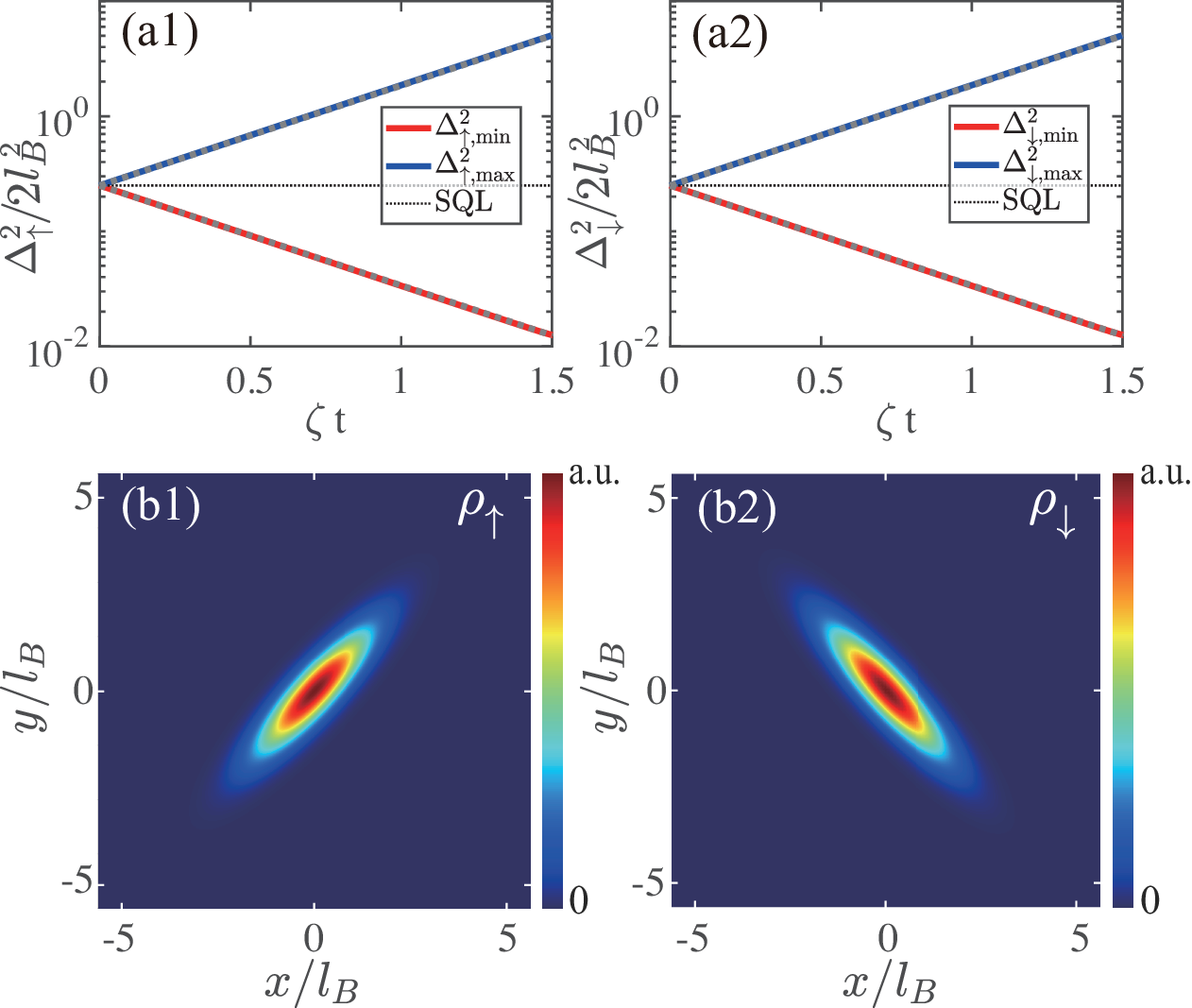}
	\caption{Squeezing dynamics for the longitudinally polarized initial states. Upper panels display the time evolution of minimum and maximum quadrature variances $\Delta_{\sigma,\min}^2$ and $\Delta_{\sigma,\max}^2$ in the guiding-center phase space, where solid lines represent numerical results and dashed lines show the analytical exponential scaling. Dotted lines show the standard quantum limit.
	Lower panels show the corresponding real-space density distributions $\rho_\sigma(\mathbf{r},t)$ at $\zeta t = 1.5$. Panels (a1)-(b1) correspond to spin-up polarized initial state, while panels (a2)-(b2) correspond to spin-down polarized initial state. In the calculation, we take $J=10\omega$, $\alpha=0.5\sqrt{\omega/m}$, $\Omega=\omega\mp\zeta$ with $\zeta = 0.00625\omega$.}
	\label{fig:single_branch_squeezing}
\end{figure}

We first consider longitudinally polarized initial states: spin-up polarized state with $\boldsymbol{\Psi}(\mathbf{r},0) = (\phi_0(\mathbf{r}),0)^\mathsf{T}$ or spin-down polarized state with $\boldsymbol{\Psi}(\mathbf{r},0) = (0,\phi_0(\mathbf{r}))^\mathsf{T}$, where 
\begin{equation}
	\phi_0(\mathbf{r})=\frac{1}{\sqrt{2 \pi}l_B}e^{-r^2/(4l_B^2)}
\end{equation}
is the ground state wave function. As discussed previously, with the rotation frequency micro-tuned to $\Omega\approx\omega\mp\zeta$  for the two cases, the system prepares perfect single-mode squeezed states. The corresponding numerical results are presented in Figs.~\ref{fig:single_branch_squeezing}(a) and (b). We take $J=10\omega$ and $\alpha=0.5\sqrt{\omega/m}$ in the calculations. Panels (a1) and (b1) show the time evolution of the quadrature fluctuations $\Delta_{\sigma,\min}^2(t)$ and $\Delta_{\sigma,\max}^2(t)$ in the $X$-$Y$ phase space for the spin-up and spin-down polarized cases, respectively. The solid lines represent numerical results, while the dashed lines correspond to the analytical exponential behavior given by Eq.~(\ref{eq:Delta_min_max}). The excellent agreement confirms the exponential squeezing and anti-squeezing dynamics.

Figure~\ref{fig:single_branch_squeezing}(b1) and (b2) display the real-space density distributions $\rho_\sigma(\mathbf{r},t)=|\psi_\sigma(\mathbf{r},t)|^2$ at $\zeta t=1.5$. For the squeezing dynamics governed by the effective Hamiltonians Eqs.~(\ref{eq:Heff_up_final}) and (\ref{eq:Heff_down_final}), the density distributions of single-mode squeezed states possess analytical expressions~\cite{fletcher2021,chen2025}
\begin{equation}
\begin{aligned}
\rho_\uparrow(\mathbf{r},t) &= \frac{e^{ -\big[1-\tanh(\zeta t)\big]\frac{(x+y)^2}{4 l_B^2}-\big[1+\tanh(\zeta t)\big]\frac{(x-y)^2}{4 l_B^2}}}{2\pi l_B^2\cosh(\zeta t)}, \\ 
\rho_\downarrow(\mathbf{r},t) &= \frac{e^{ -\big[1+\tanh(\zeta t)\big]\frac{(x+y)^2}{4 l_B^2}-\big[1-\tanh(\zeta t)\big]\frac{(x-y)^2}{4 l_B^2}}}{2\pi l_B^2\cosh(\zeta t)} .
\end{aligned}
\label{eq:density_both}
\end{equation}
Both distributions are 2D anisotropic Gaussian, with the density stretched along the $\pi/4$ direction for spin-up and along the $-\pi/4$ direction for spin-down, respectively. The numerical density profiles shown in Figs.~\ref{fig:single_branch_squeezing}(b1) and (b2) are in excellent agreement with the analytical predictions.

\begin{figure}[t]
	\centering
	\includegraphics[width=\linewidth]{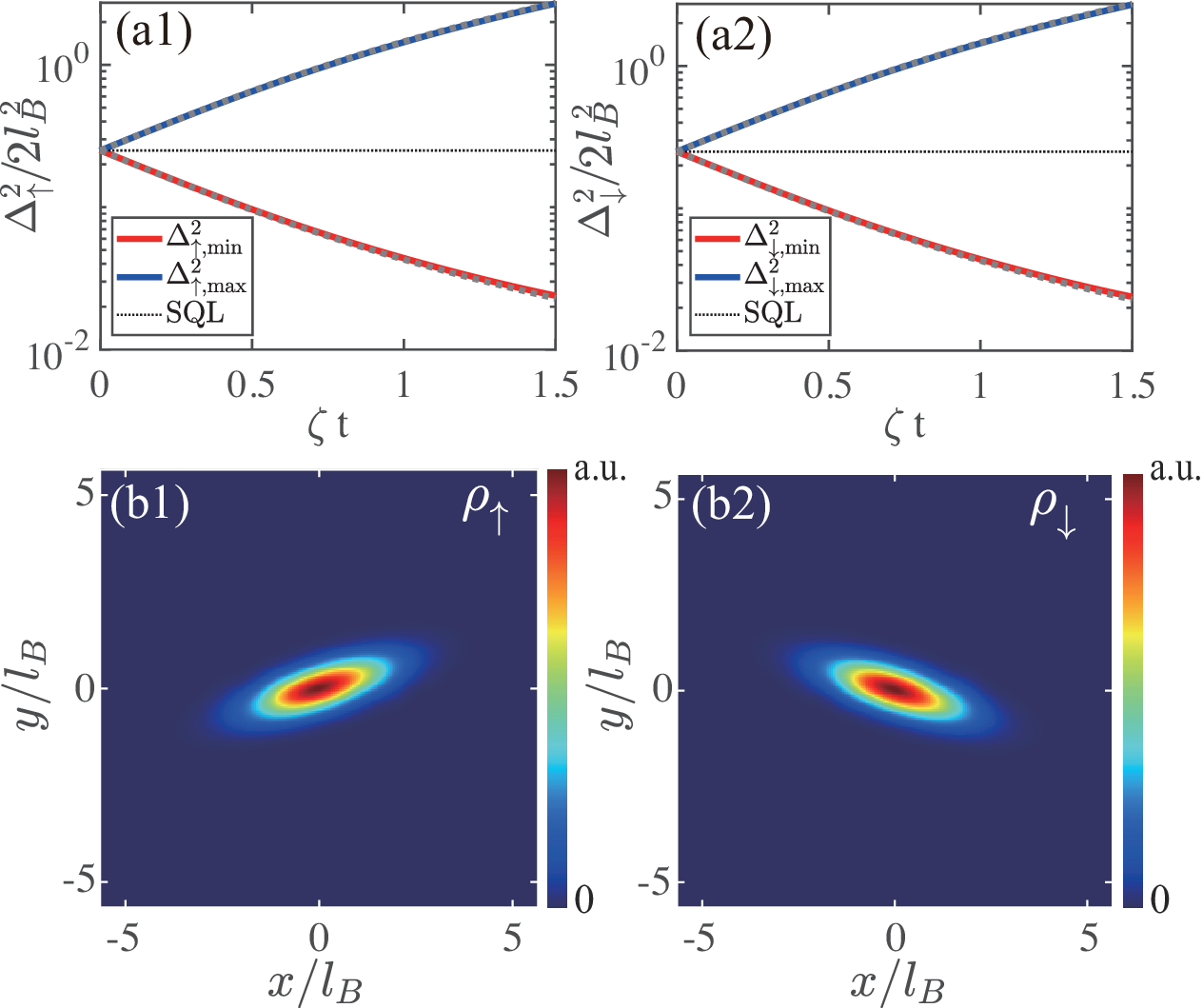}
	\caption{Squeezing dynamics for the transversely polarized initial state at critical rotation $\Omega=\omega$. Upper panels show the time evolution of quadrature variances $\Delta_{\sigma,\min}^2$ and $\Delta_{\sigma,\max}^2$, where solid lines represent numerical results and dashed lines show the analytical sheared squeezing Eq.~(\ref{eq:shear_squeezing}). Lower panels display the density distributions $\rho_\sigma(\mathbf{r},t)$ at $\zeta t=1.5$. Panels (a1)-(b1) and (a2)-(b2) correspond to spin-up and spin-down components, respectively. The calculation parameters are same as those used in Fig.~\ref{fig:single_branch_squeezing}.}
		\label{fig:equal_superposition}
\end{figure}

Next, we consider the transversely polarized initial state $\boldsymbol{\Psi}(\mathbf{r},0)=\frac{1}{\sqrt{2}}(\phi_0(\mathbf{r}),\phi_0(\mathbf{r}))^\mathsf{T}$, i.e., an equal-weight superposition of spin-up and spin-down components. The dynamical results are shown in Fig.~\ref{fig:equal_superposition}, where $\Omega = \omega$ is fixed. As predicted in the previous subsection, this leads to the sheared squeezing dynamics due to the residual diagonal strengths $\mp\zeta$ for the two spin components. Panels (a1) and (a2) show the phase-space fluctuation dynamics for both components. The solid lines represent numerical results, while the dashed lines correspond to the analytical sheared squeezing behavior given by Eq.~(\ref{eq:shear_squeezing}). The numerical and analytical results are in excellent agreement.
Compared to the polarized cases in Fig.~\ref{fig:single_branch_squeezing}, the exponential behavior is still evident at short times, 
but at longer times $\zeta t\sim 1$, the residual diagonal terms become relevant, causing the quadrature fluctuations to deviate from the ideal exponential scaling. 
At $\zeta t=1.5$, as shown in panels (b1) and (b2), the density distributions also exhibit anisotropic Gaussian patterns, but the stretch directions are deviated from $\pm\pi/4$.

\section{Two-Mode Geometric Squeezing}
\label{sec:two_mode_squeezing}

In the above, we have shown that weak SOC induces single-mode squeezing in the two LLL spin subspaces. Here, we explain how these two single-mode squeezing channels can be converted into a two-mode geometrically squeezed state by a linear mode mixing.
The key idea is that, according to the Bloch-Messiah reduction~\cite{braunstein2005,weedbrook2012}, any multimode Gaussian unitary can be decomposed into passive linear mixing and single-mode squeezing. In our cold-atom setting, spin mixing is implemented by a $\pi/2$ spin rotation ($\pi/2$ pulse), which is formally equivalent to a 50:50 linear mixer in quantum optics.

\begin{figure}[t]
	\centering
	\includegraphics[width=0.40\textwidth]{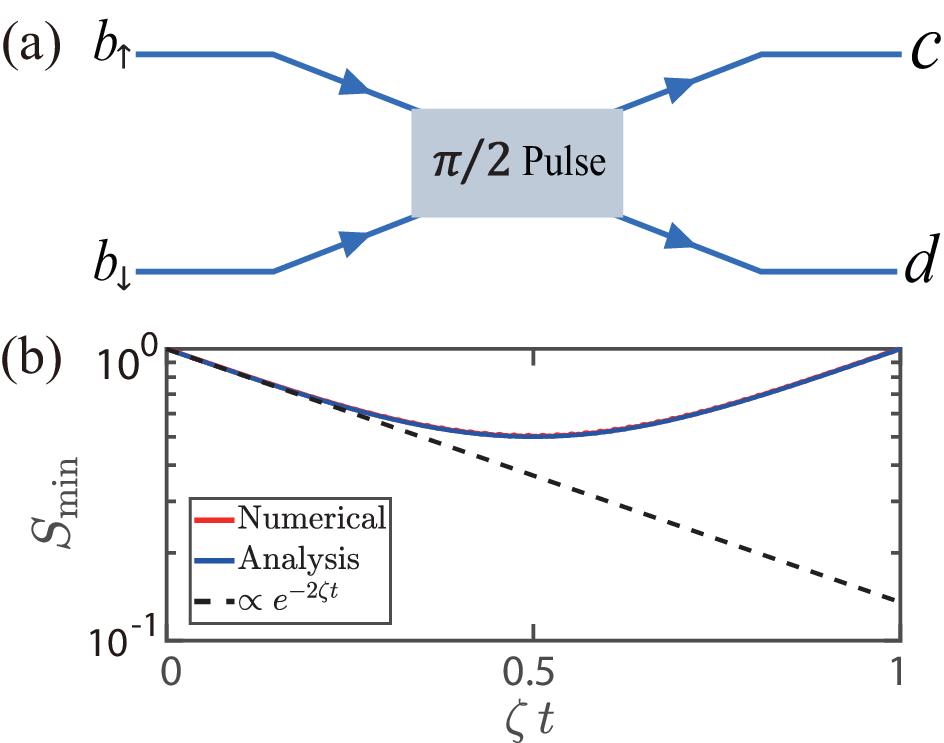}
	\caption{(a) Schematic illustration of the generation of two-mode geometric squeezing using a mode transformation. The $\pi/2$ spin rotation maps the input modes $(b_\uparrow, b_\downarrow)$ to the output modes $(c, d)$. (b) Time evolution of the minimum joint noise $S_{\min}(t)$. The red and blue solid lines denote the numerical and analytical results, respectively. The black dashed line shows the ideal exponential scaling $\propto e^{-2\zeta t}$ for a standard two-mode squeezing Hamiltonian. All parameters are the same as those used in Fig.~\ref{fig:equal_superposition}.
	}
	\label{fig4}
\end{figure}

We consider the effective Hamiltonian of LLL within guiding-center subspaces [according to Eqs.~(\ref{eq:Heff_up_final0}) and (\ref{eq:Heff_down_final0}) with constant terms being neglected]
\begin{equation}
	H_{\rm eff}^{\rm LLL} = \frac{\zeta}{2}\big[(b_\uparrow^\dagger)^2 +  b_\uparrow^2\big]
	- \zeta\, b_\uparrow^\dagger  b_\uparrow
	- \frac{\zeta}{2}\big[( b_\downarrow^\dagger)^2 +  b_\downarrow^2\big]
	+ \zeta\, b_\downarrow^\dagger  b_\downarrow ,
	\label{eq:H_eff_spin}
\end{equation}where the quadratic terms have opposite signs for $b_\uparrow$ and $b_\downarrow$, corresponding to orthogonal single-mode squeezing directions as mentioned before. The relation between the modes before and after the $\pi/2$ spin rotation [illustrated in Fig.~\ref{fig4}(a)] are
$c = (b_\downarrow + b_\uparrow)/\sqrt{2}$ and $d = (b_\downarrow - b_\uparrow)/\sqrt{2}$.
Rewriting Eq.~\eqref{eq:H_eff_spin} in terms of $c$ and $d$ yields 
\begin{equation}
	H_{\rm eff}^{\rm LLL} = -\zeta (c^\dagger d^\dagger + c d) + \zeta(c^\dagger d + d^\dagger c),
	\label{eq:H_eff_cd}
\end{equation}
where the first two terms form the standard two-mode squeezing Hamiltonian, whereas 
the last two terms are mode-exchange terms which originate from the residual diagonal contributions $\propto b_\sigma^\dagger b_\sigma$ in Eq.~\eqref{eq:H_eff_spin}. 
The mode-exchange term is of the same order as the two-mode squeezing term and therefore qualitatively modifies the finite-time dynamics.

It is worth mentioning that the connection between single-mode and two-mode squeezing can be also understood from the perspective of $\mathfrak{su}(1,1)$ algebra. The $\mathfrak{su}(1,1)$ generators satisfy the commutation relations $[\hat{K}_0, \hat{K}_\pm] = \pm \hat{K}_\pm$ and $[\hat{K}_+, \hat{K}_-] = -2\hat{K}_0$.
For the single-mode squeezing Hamiltonians [Eqs.~(\ref{eq:Heff_up_final0}) and (\ref{eq:Heff_down_final0})], the generators are $\hat{K}_0^{(\sigma)} = (b_\sigma^\dagger b_\sigma + \frac{1}{2})/2$, $\hat{K}_+^{(\sigma)} = (b_\sigma^\dagger)^2/2$, and $\hat{K}_-^{(\sigma)} = b_\sigma^2/2$ with $\sigma \in \{\uparrow, \downarrow\}$.
For the two-mode squeezing part in Eq.~(\ref{eq:H_eff_cd}), the generators become $\hat{K}_0 = (c^\dagger c + d^\dagger d + 1)/2$, $\hat{K}_+ = c^\dagger d^\dagger$, and $\hat{K}_- = c d$.
The two squeezing scenarios correspond to different representations of the $\mathfrak{su}(1,1)$ algebra, which are connected by the $\pi/2$ transformation.

We quantify the two-mode squeezing by the joint-noise factor~\cite{duan2000,simon2000}

\begin{equation}
	S(\theta,t)
	= \frac{1}{l_B^2}
	\left[\big\langle (\Delta \hat{u}_\theta)^2 \big\rangle
	+ \big\langle (\Delta \hat{v}_\theta)^2 \big\rangle\right],
	\label{eq:S_definition}
\end{equation}
where $u_\theta = \big(X_c(\theta)+X_d(\theta)\big)/\sqrt{2}$ and $v_\theta = \big(Y_c(\theta)-Y_d(\theta)\big)/\sqrt{2}$ are the joint operators. Here, we have defined the quadrature operators $X_{\beta\in(c,d)} = l_B (\beta + \beta^\dagger)/\sqrt{2}$ and $Y_\beta = l_B (\beta - \beta^\dagger)/(-i\sqrt{2})$,  and $X_\beta(\theta)\equiv X_\beta\cos\theta+Y_\beta\sin\theta$ are the rotated quadratures, with $\beta\in\{c,d\}$ and $\theta$ being the measurement angle.
Note that the vacuum state has $S=1$. The cases with $S<1$ indicate the two-mode squeezing.

For the dynamics generated by Eq.~\eqref{eq:H_eff_cd}, one obtains
\begin{equation}
	S(\theta,t)
	= 1 - 2\zeta t\sin2\theta + 2\zeta^2t^2.
	\label{eq:Duan_theta_general}
\end{equation}
Taking the derivative of Eq.~\eqref{eq:Duan_theta_general} with respect to $\theta$, the optimal joint measurement angle is determined by $\theta_{\mathrm{opt}} = \pi/4$(mod $\pi$).
Substituting $\theta_{\mathrm{opt}}$ back into Eq.~\eqref{eq:Duan_theta_general}, we obtain the minimum factor
\begin{equation}
	S_{\min}(t)
	= 1 - 2\zeta t + 2\zeta^2t^2,
	\label{eq:Duan_min_square}
\end{equation}
where the joint-noise is reduced below the vacuum level only within the finite time window $0<\zeta t<1$. The strongest two-mode squeezing occurs at $\zeta t=1/2$, where $S_{\min}=1/2$. 
Therefore, in contrast to an ideal standard two-mode squeezing Hamiltonian, for which $S_{\min}(t)\propto e^{-2\zeta t}$ decreases monotonically, the present full effective Hamiltonian produces transient two-mode squeezing. 
This transient behavior is caused by the number-conserving mode-exchange term in Eq.~\eqref{eq:H_eff_cd}. In Fig.~\ref{fig4}(b), the numerical result agrees well with the analytical expression in Eq.~\eqref{eq:Duan_min_square}.  

\section{Influence of Interatomic Interactions}
\label{sec:discussion}
\begin{figure}[t]
	\centering
	\includegraphics[width=0.40\textwidth]{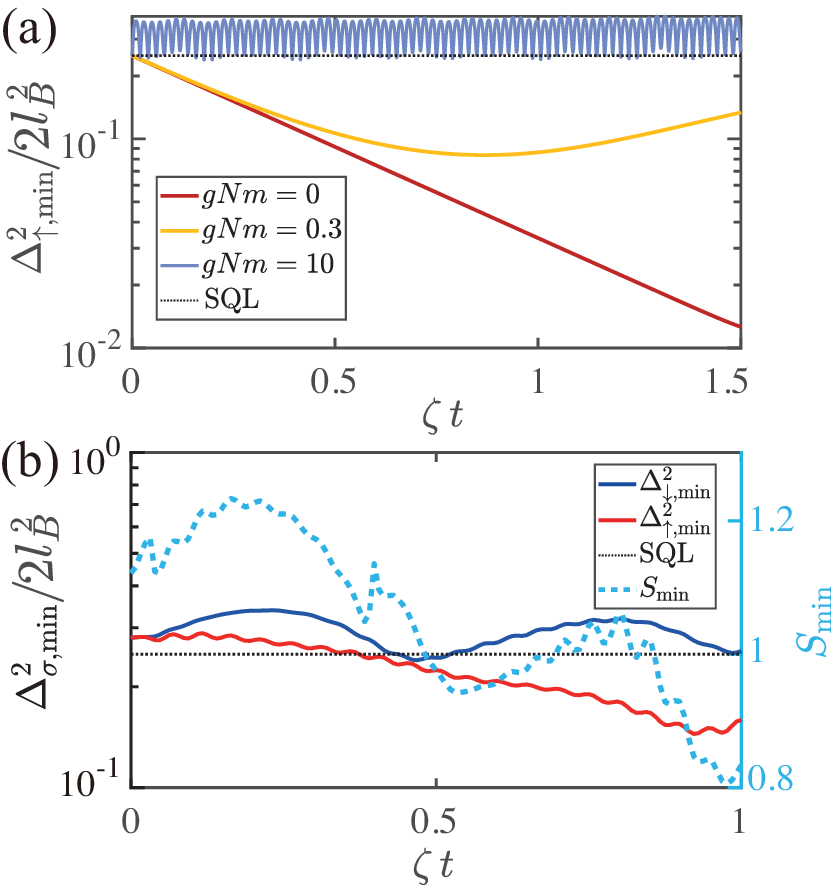}
	\caption{Squeezing dynamics of interacting BECs. (a) Time evolution of $\Delta_{\uparrow,\min}^2$ for the spin-up component at $J = 10 \omega$, $\alpha=0.5\sqrt{\omega/m}$ and $\Omega=\omega-\zeta$ with $\zeta = 0.00625\omega$. Different lines correspond to different interaction strengths $g$. (b) Time evolution of $\Delta_{\sigma,\min}^2$ (left axis) for both spin components and the minimum joint noise $S_{\min}$ (right axis) for at fixed interaction strength $gNm=10$, $\alpha = 2.5\sqrt{\omega/m}$ and $\Omega=0.6\omega$. The initial state is the longitudinally polarized state for panel (a) and the transversely polarized state for panel (b).}
		\label{fig5}
\end{figure}

In the preceding sections, we have discussed the physical picture of SOC-induced geometric squeezing based on the single-particle Hamiltonian $H$ [Eq.~(\ref{eq:H_total})]. Now, we investigate the influence of interatomic interactions on the squeezing dynamics. 
To this end, we numerically solve the time-dependent Gross-Pitaevskii (GP) equation, which is the mean-field description of BECs with two-body s-wave collisions. The GP equation for the spin-1/2 BEC is in the well-known form~\cite{lin2011,ho2011,li2012}
\begin{equation}
	i \frac{\partial}{\partial t} \boldsymbol{\Psi}(\mathbf{r},t) = \left( H + \mathcal{G} \right) \boldsymbol{\Psi}(\mathbf{r},t),
\end{equation}
where
\begin{equation}
	\mathcal{G} = \begin{pmatrix}
		g_{\uparrow\uparrow}|\psi_{\uparrow}|^2 + g_{\uparrow\downarrow}|\psi_{\downarrow}|^2 & 0 \\
		0 & g_{\downarrow\downarrow}|\psi_{\downarrow}|^2 + g_{\uparrow\downarrow}|\psi_{\uparrow}|^2
	\end{pmatrix}
\end{equation}
with $g_{\sigma\sigma}$ and $g_{\uparrow\downarrow}$ denoting the intra-spin and inter-spin interactions, respectively.
In our following calculations, we simply take $g_{\uparrow\uparrow} = g_{\downarrow\downarrow} = g_{\uparrow\downarrow} = g$. 
This condition can be naturally satisfied in the most commonly used atomic species, such as $^{23}$Na and $^{87}$Rb, where the difference between $g_{\sigma\sigma}$ and $g_{\uparrow\downarrow}$ is less than a few percent~\cite{kawaguchi2012,lin2011}. Our numerical results are presented in Fig.~\ref{fig5}.
The reduced two-dimensional s-wave interaction is given by $g=\sqrt{8\pi\omega_z/m}\,a_s$ \cite{chen2025,chen2025dyn}, where $a_s$ is the three-dimensional $s$-wave scattering length and $\omega_z$ is the trapping frequency along the z-direction. Hence, the dimensionless interaction strength used in the simulations is $gNm=N\sqrt{8\pi m\omega_z}\,a_s$. For a quasi-two-dimensional $^{23}$Na condensate with $a_s\simeq52a_0$ \cite{kawaguchi2012}, $\omega_\perp/2\pi\simeq90$ Hz, and $\omega_z/\omega_\perp\simeq\sqrt{8}$, this gives $gNm\simeq0.010N$. Particularly for $N = 10^3$ atoms, $gNm\approx 10$. 
In our calculations, we gradually increase the dimensionless interaction strength $gNm$ from the non-interacting limit ($gNm = 0$), and examine the changes in the squeezing dynamics.

Fig.~\ref{fig5}(a) displays the minimum variance of the guiding-center mode for the spin-up component, i.e., $\Delta_{\uparrow,\min}^2$, with different curves corresponding to varying interaction strength $g$. 
Here, the initial state is set to be the spin-up polarized state, i.e., $\boldsymbol{\Psi}(\mathbf{r},0) = (\phi_\mathrm{g}(\mathbf{r}),0)^\mathsf{T}$, in which $\phi_\mathrm{g}(\mathbf{r})$ is the isotropic ground-state wave function ($\alpha = 0$) for a given interaction strength $g$. 
All other parameters, except for $g$, are the same as those in Fig.~\ref{fig:single_branch_squeezing}(a1).
Particularly, $g=0$ exactly corresponds to the non-interacting case whose dynamics has already been shown in Fig.~\ref{fig:single_branch_squeezing}(a1). 

As clearly shown in Fig.~\ref{fig5}(a), the interaction $g$ suppresses the generation of geometric squeezing. 
For $gNm = 0.3$, the short-time dynamics still exhibit an exponential decay (similar to the non-interacting case), while the long-time behavior deviates substantially from the exponential trend and eventually exhibits an upward drift. When $gNm = 10$, the squeezing effect completely vanishes, with $\Delta_{\uparrow,\min}^2$ exhibiting weak periodic oscillations around its initial value. 
Similar oscillatory phenomena also occur as the initial state is prepared in the spin-down polarized state or the transversely polarized state.
In fact, similar periodic oscillatory behavior has already been observed in scalar BECs~\cite{chen2025dyn}, where the underlying mechanism is that interactions modify the BEC stability, and the oscillation corresponds to a certain collective excitation near the equilibrium. 

We further find that by increasing $\zeta$ (namely increasing $\alpha$) and reducing $\Omega$ away from unity, we can achieve a stronger squeezing effect. Specifically, we fix $gNm = 10$, choose $\alpha = 2.5\sqrt{\omega/m}$ and $\Omega = 0.6\omega$, and then propagate the GP equation from the transversely polarized initial state $\boldsymbol{\Psi}(\mathbf{r},0) = (\phi_\mathrm{g}(\mathbf{r}),\phi_\mathrm{g}(\mathbf{r}))^\mathsf{T}/\sqrt{2}$. Fig.~\ref{fig5}(b) shows the minimal quadrature fluctuations $\Delta_{\sigma,\min}^2$ for both the spin-up and spin-down components. In addition, we also display the minimum joint-noise factor $S_{\min}$ (quantified on the right axis), which corresponds to the two-mode squeezing protocol illustrated in Fig.~\ref{fig4}(a). As can be seen from the figure, the spin-up fluctuation $\Delta_{\uparrow,\min}^2$ exhibits an exponential-like squeezing behavior, whereas the spin-down fluctuation $\Delta_{\downarrow,\min}^2$ shows periodic oscillations without squeezing. Benefiting from the former, two-mode squeezing begins to emerge at $\zeta t \gtrsim 0.8$ (manifested by $S_{\min} < 1$). Additionally, note that, based on our limited numerical explorations, we have not yet found a parameter regime where both spin components simultaneously exhibit exponential-like squeezing behavior.

\section{Discussion}
\label{sec:observation}

We first discuss the experimental realization of our proposal.
Unlike conventional static SOC configurations, the rotating-frame Hamiltonian $H$ [Eq.~(\ref{eq:H_total})] requires the Raman recoil direction to co-rotate with the trap.
A viable scheme, originally proposed in Ref.~\cite{Radic2011}, is to let the Raman recoil direction to co-rotate with the trap, then the laboratory-frame Hamiltonian takes the form
\begin{equation}
\begin{aligned}
    H_{\rm lab} =\ & 
    \left[ \frac{\mathbf{p}^2}{2m} + V_0\big(R^{-1}(t)\mathbf{r}\big) \right] \otimes \mathbb{I} 
    - J \sigma_z \\
    & +\, \alpha \big( p_x \cos \Omega t + p_y \sin \Omega t \big) \sigma_x,
\end{aligned}
    \label{eq:H_lab}
\end{equation}
which, under the standard rotation transformation $U(t)=e^{i\Omega t L_z}$, reduces exactly to Eq.~(\ref{eq:H_total}).
Since the recoil direction in a Raman scheme is set by the relative geometry of the laser beams, this co-rotation could be implemented, for example, via acousto-optic deflectors or rotating optical elements; however, such schemes require additional engineering effort compared to a static beam geometry.

We now outline how the guiding-center quadrature variances $\Delta_\sigma^2(\theta,t)$ and the principal extrema $\Delta_{\sigma,\min/\max}^2(t)$ can be reconstructed from experimentally accessible observables. The key observation is that the real-space coordinates split into guiding-center and cyclotron parts, $x=X_\sigma+\xi_\sigma$ and $y=Y_\sigma+\eta_\sigma$, so the real-space projection along an arbitrary direction $\theta$ is the sum $r_\theta^{(\sigma)}=X_\sigma(\theta)+P_{\theta,{\rm cyc}}^{(\sigma)}$ of the guiding-center quadrature $X_\sigma(\theta)=X_\sigma\cos\theta+Y_\sigma\sin\theta$ defined in Eq.~(\ref{eq:Delta_theta}) and the corresponding cyclotron quadrature $P_{\theta,{\rm cyc}}^{(\sigma)}=\xi_\sigma\cos\theta+\eta_\sigma\sin\theta.$ 
Since the two sectors are statistically independent, the real-space variance $W_\sigma^2(\theta,t)\equiv\langle r_\theta^{(\sigma)\,2}\rangle-\langle r_\theta^{(\sigma)}\rangle^2$ along the direction $\theta$ satisfies
\begin{equation}
	W_\sigma^2(\theta,t)=\Delta_\sigma^2(\theta,t)+\langle P_{\theta,{\rm cyc}}^{(\sigma)\,2}\rangle.
	\label{eq:W_decomposition}
\end{equation}

\begin{figure}[t]
	\centering
	\includegraphics[width=0.40\textwidth]{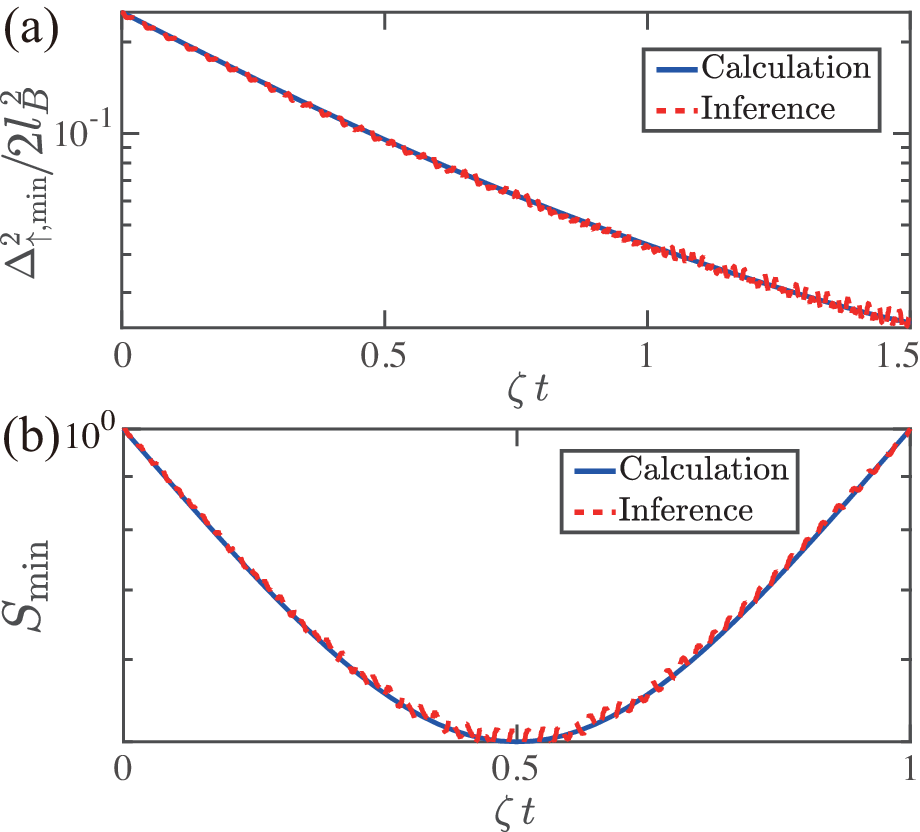}
	\caption{The inference of single-mode (a) and two-mode (b) geometric squeezing by density distributions. (a) Time evolution of the minimum guiding-center quadrature variance $\Delta_{\uparrow,\min}^2(t)$ obtained by calculation (solid line) and by the inferred observation (dashed line) from Eq.~\eqref{eq:Delta_from_W}. (b) Time evolution of the minimum joint-noise factor $S_{\min}(t)$ obtained by calculation (solid line) and by the inferred observation (dashed line) from Eq.~\eqref{eq:S_from_W}.
	}
	\label{fig6}
\end{figure}

For a Gaussian state in the LLL regime—which corresponds exactly to the non-interacting BEC considered in Secs.~\ref{sec:aniso_squeezing}-\ref{sec:two_mode_squeezing}—the cyclotron mode remains in its zero-point vacuum and contributes the $\theta$-independent floor $\langle P_{\theta,{\rm cyc}}^{(\sigma)\,2}\rangle=l_B^2/2$, such that
\begin{equation}
	\Delta_\sigma^2(\theta,t)=W_\sigma^2(\theta,t)-\frac{l_B^2}{2}.
	\label{eq:Delta_from_W}
\end{equation}
Experimentally, $W_\sigma^2(\theta,t)$ can be obtained from the spin-resolved in-situ density profile $\rho_\sigma(\mathbf r,t)$ via the second moments $C_{ij}^{(\sigma)}=\langle (r_i-\bar r_{i,\sigma})(r_j-\bar r_{j,\sigma})\rangle_\sigma$ ($i,j\in\{x,y\}$), giving
\begin{equation}
	W_\sigma^2(\theta,t)=C_{xx}^{(\sigma)}\cos^2\theta+2C_{xy}^{(\sigma)}\sin\theta\cos\theta+C_{yy}^{(\sigma)}\sin^2\theta.
	\label{eq:W_theta}
\end{equation}
The principal extrema $\Delta_{\sigma,\min/\max}^2(t)$ are obtained by diagonalizing $C^{(\sigma)}$ and subtracting $l_B^2/2$, and the squeezing-axis orientation follows from $\tan(2\theta_\sigma)={2C_{xy}^{(\sigma)}}/({C_{xx}^{(\sigma)}-C_{yy}^{(\sigma)}})$.
Moreover, using the mode transformation $c=(b_\downarrow+b_\uparrow)/\sqrt{2}$ and $d=(b_\downarrow-b_\uparrow)/\sqrt{2}$, the joint-noise factor $S(\theta,t)$ defined in Eq.~(\ref{eq:S_definition}) can be re-expressed as
\begin{align}
	S(\theta,t) 
	&= \frac{1}{l_B^2} \left[ \Delta_\downarrow^2(\theta,t) + \Delta_\uparrow^2\left(\theta+\frac{\pi}{2}, t\right) \right] \notag \\
	&= \frac{1}{l_B^2} \left[ W_\downarrow^2(\theta,t) + W_\uparrow^2\left(\theta+\frac{\pi}{2}, t\right) - l_B^2 \right],
	\label{eq:S_from_W}
\end{align}
which is also fully determined by the spin-resolved density distributions.

For more general non-Gaussian states, particularly those with significant Landau-level mixing or strong interactions, the reconstruction becomes considerably more involved: the cyclotron mode may no longer remain in its zero-point vacuum, and the guiding-center and cyclotron sectors can develop nontrivial correlations, such that neither Eq.~(\ref{eq:Delta_from_W}) nor the decomposition Eq.~(\ref{eq:W_decomposition}) remains valid.
Therefore, developing measurement protocols for the guiding-center fluctuations of non-Gaussian states is an interesting problem that deserves further investigation.

We numerically verify the density-based inference schemes in Eqs.~\eqref{eq:Delta_from_W} and \eqref{eq:S_from_W}, and present the inferred results as dashed lines in Figs.~\ref{fig6}(a) and \ref{fig6}(b), respectively. Panels (a) and (b) correspond to the dynamics of geometric squeezing starting from a transversely polarized initial state. Specifically, panel (a) corresponds to the case discussed in Fig.~\ref{fig:equal_superposition}(a1), while panel (b) corresponds to the case in Fig.~\ref{fig4}(b). The solid lines in both panels represent the results obtained by numerical calculation, which have been previously displayed in Fig.~\ref{fig:equal_superposition}(a1) and Fig.~\ref{fig4}(b). The results inferred from the density distribution match the calculated results.

\section{CONCLUSION}
\label{sec:conclusion}
We have demonstrated that, in a rotating pseudospin-1/2 Bose-Einstein condensate, a weak Raman spin-orbit coupling is capable of inducing single-mode geometric squeezing within each spin component. The underlying mechanism originates from spin-dependent momentum transfer, which drives effective two-phonon processes within the lowest Landau-level manifold. Furthermore, these two independent squeezing channels can be converted into two-mode squeezing via a passive $\pi/2$ spin rotation.
We further showed that, within the Gross-Pitaevskii mean-field framework, many-body interactions suppress the dynamical squeezing. 
By enhancing the spin-orbit coupling strength and adjusting the rotation frequency, we can restore the geometric squeezing in interacting BECs.
Our results provide a theoretical foundation for experimentally realizing and manipulating rich single-mode and two-mode geometrically squeezed states in spinor quantum gases.
Looking forward, laser-induced synthetic gauge fields — which generate an effective magnetic field analogous to the Coriolis force in a rotating frame~\cite{Lin2009PRL,Lin2009Nature} — offer a complementary platform for realizing Landau levels without mechanical rotation; extending the present geometric-squeezing mechanism to such synthetic-field systems is an interesting direction for future work.

\begin{acknowledgments}
We acknowledge support from the NSF of China (Grant Nos. 12574296, U25A20198, 12574299, 12474266), the Sanjin Talent Program of Shanxi Province, and the fund for the Shanxi 1331 Project. 
\end{acknowledgments}

\appendix

\section{Squeezing Hamiltonian via Anisotropic Potential}
\label{app:eq9_validity}

We present the derivation of the effective squeezing Hamiltonian~\eqref{eq:h0_anisotropic} from the rotating BEC Hamiltonian with weak anisotropic confinement. Using the Baker--Campbell--Hausdorff formula, one finds that $G$ leaves the spatial coordinates invariant,
\begin{equation}
	GxG^\dagger = x, \qquad GyG^\dagger = y,
	\label{eq:app_G_coords}
\end{equation}
while the canonical momenta are shifted linearly in $\kappa$,
\begin{equation}
	Gp_xG^\dagger = p_x + \kappa m\omega\, y, \qquad
	Gp_yG^\dagger = p_y + \kappa m\omega\, x.
	\label{eq:app_G_momenta}
\end{equation}
Substituting Eq.~\eqref{eq:app_G_momenta} into the full Hamiltonian $\mathbf{p}^2/(2m)+V(\mathbf{r})-\Omega L_z$ [with $V(\mathbf{r})$ given in Eq.~\eqref{eq:V_aniso}] and choosing $\kappa=\varepsilon\omega/(2\Omega)$ to cancel the anisotropic term, the transformed Hamiltonian $\tilde{h}_0=Gh_0G^\dagger$ takes the exact form
\begin{align}
	\tilde{h}_0 &= \frac{\mathbf{p}^2}{2m}
	+\frac{m\omega^2(1+\kappa^2)}{2}(x^2+y^2)
	\nonumber\\
	&\quad -\Omega L_z +\kappa\omega(xp_y+yp_x).
	\label{eq:app_h_tilde}
\end{align}
The squeezing physics is now encoded in the cross term $\kappa\omega(xp_y+yp_x)$.

To express Eq.~\eqref{eq:app_h_tilde} in terms of ladder operators, we introduce scaled canonical variables that absorb the renormalized trap frequency $\omega\sqrt{1+\kappa^2}$,
\begin{align}
	\tilde{x} &= (1+\kappa^2)^{1/4}x, &
	\tilde{p}_x &= (1+\kappa^2)^{-1/4}p_x, \notag\\
	\tilde{y} &= (1+\kappa^2)^{1/4}y, &
	\tilde{p}_y &= (1+\kappa^2)^{-1/4}p_y,
	\label{eq:app_scaled}
\end{align}
and the corresponding scaled cyclotron and guiding-center quadratures,
\begin{align}
	\tilde{\xi} &= \frac{\tilde{x}}{2}-\frac{\tilde{p}_y}{2m\omega}, &
	\tilde{\eta} &= \frac{\tilde{y}}{2}+\frac{\tilde{p}_x}{2m\omega}, \notag\\
	\tilde{X} &= \frac{\tilde{x}}{2}+\frac{\tilde{p}_y}{2m\omega}, &
	\tilde{Y} &= \frac{\tilde{y}}{2}-\frac{\tilde{p}_x}{2m\omega},
	\label{eq:app_quadratures}
\end{align}
together with the associated ladder operators,
\begin{align}
	\tilde{a} &= \sqrt{m\omega}\,(\tilde{\xi}+i\tilde{\eta}), &
	\tilde{a}^\dagger &= \sqrt{m\omega}\,(\tilde{\xi}-i\tilde{\eta}), \notag\\
	\tilde{b} &= \sqrt{m\omega}\,(\tilde{X}-i\tilde{Y}), &
	\tilde{b}^\dagger &= \sqrt{m\omega}\,(\tilde{X}+i\tilde{Y}).
	\label{eq:app_ladder}
\end{align}
After straightforward algebra, the transformed Hamiltonian~\eqref{eq:app_h_tilde} expressed in the scaled basis reads
\begin{align}
	\tilde{h}_0={}&
	(\lambda\omega+\Omega)\!\left(\tilde{a}^\dagger\tilde{a}+\tfrac{1}{2}\right)
	+(\lambda\omega-\Omega)\!\left(\tilde{b}^\dagger\tilde{b}+\tfrac{1}{2}\right)
	\nonumber\\
	&-\frac{\kappa\omega}{2}\!\left(
	\tilde{a}^{\dagger 2}+\tilde{a}^2
	-\tilde{b}^{\dagger 2}-\tilde{b}^2
	\right),
	\label{eq:app_h_scaled}
\end{align}
with $\lambda=\sqrt{1+\kappa^2}$. This is the squeezed-mode representation of the transformed Hamiltonian.

Expressing $(\tilde{a},\tilde{b})$ back in terms of the original cyclotron and guiding-center operators $(a,b)$ defined in Sec.~\ref{sec:aniso_squeezing}, the result is
\begin{align}
	\tilde{h}_0={}&
	\left[\Omega+\omega\!\left(1+\frac{\kappa^2}{2}\right)\right]a^\dagger a
	+\left[-\Omega+\omega\!\left(1+\frac{\kappa^2}{2}\right)\right]b^\dagger b
	\nonumber\\
	&+\frac{\omega\kappa^2}{2}(a^\dagger b^\dagger+ab)
	-\frac{\kappa\omega}{2}\!\left(
	a^{\dagger 2}+a^2-b^{\dagger 2}-b^2
	\right)
	\nonumber\\
	&+\text{const.}
	\label{eq:app_h_exact}
\end{align}
The hierarchy of terms in Eq.~\eqref{eq:app_h_exact} is transparent. The dominant single-mode squeezing sector $-(\kappa\omega/2)(a^{\dagger 2}+a^2-b^{\dagger 2}-b^2)$, with $\kappa\omega=\zeta$, is of order $O(\kappa)\sim O(\varepsilon)$. 
Neglecting the $O(\kappa^2)\sim O(\varepsilon^2)$ terms in Eq.~\eqref{eq:app_h_exact} yields Eq.~\eqref{eq:h0_anisotropic}.

\end{document}